\documentclass{aa}
\usepackage{graphicx}
\usepackage{epsf}
\usepackage{psfig}
\usepackage{natbib}
\usepackage{latexsym}
\usepackage{amsmath}
\usepackage{amssymb}
\usepackage{amstext}
\usepackage{color}
\usepackage{psfrag}

\def\d{$^\circ$}
\def\m{$^\prime$}
\def\s{$^{\prime\prime}$}
\def\hh{$^{\mathrm h}$}
\def\mm{$^{\mathrm m}$}
\def\ss{$^{\mathrm s}$}
\def\cm3{cm$^{-3}$}

\def\cxou{CXOU J174722.8-280915}

\begin{document}
\title { High resolution radio study of the Pulsar Wind Nebula within
the Supernova Remnant G0.9+0.1}
\author{G. Dubner\inst{1}
       \and E. Giacani\inst{1}
       \and A. Decourchelle\inst{2}
        }

\titlerunning{Radio observations of the PWN in  G0.9+0.1}
\authorrunning{Dubner et al.}

    \offprints{G. Dubner}

\institute{Instituto de Astronom\'{\i}a y  F\'{\i}sica del Espacio (IAFE),
CC 67, Suc. 28, 1428 Buenos Aires, Argentina\\
             \email{gdubner@iafe.uba.ar, egiacani@iafe.uba.ar
           \thanks{G.Dubner and E. Giacani are members of the 
Carrera del Investigador Cient\'{\i}fico of CONICET, Argentina}}
\and
         Service d'Astrophysique, Orme des Merisiers, CE-Saclay,
         91191 Gif-sur-Yvette, Cedex, France\\
         \email{anne.decourchelle@cea.fr}
 }
\date{Submitted:~~~~~~~~~~; Accepted: ~~~~ ~~~~~~}

\abstract{
{\sl Aims.} We have conducted a study in radio wavelengths and in
X-rays of the 
pulsar wind nebula (PWN) in the  supernova remnant (SNR) G0.9+0.1   
with the goal of investigating in detail its morphology and to
accurately determine its characteristic parameters.

{\sl Method.} To carry out this research we have observed the PWN at
$\lambda$ 3.6 and 6 cm using the Australia Telescope Compact Array
(ATCA) and combined these data with existing multiconfiguration 
VLA data and
single dish observations in order to recover information at all
spatial scales. We have also reprocessed VLA archival data at
$\lambda$ 20 cm. From all these observational data we have produced
high-fidelity images at the three radio frequencies with angular resolution 
better than 3\s. The radio data were compared to X-ray data 
obtained with {\it Chandra} and in two different observing runs with 
{\it XMM-Newton}.

{\sl Results.}  The new observations revealed that the 
morphology  and symmetry suggested by {\it Chandra} observations
(torus and jet-like features) are basically preserved in the radio
range in spite of the rich structure observed in the  
radio emission of this PWN, including 
several arcs, bright knots, extensions and filaments. The reprocessed X-ray images show for the first time that the X-ray
plasma fills almost the same volume as the radio PWN. Notably the
X-ray maximum does not coincide with the radio maximum and the neutron
star candidate \cxou~ lies within a small depression in the
radio emission.
From the new
radio data we have refined the flux density estimates, obtaining
S$_{\rm PWN}\sim 1.57$ Jy, almost constant between $\lambda$ 3.6
and $\lambda$ 20 cm. For the whole SNR (compact core and shell),
a flux density S$_{\rm 20 cm}$= 11.5 Jy was estimated.  
Based on  the new and
the existing $\lambda$ 90 cm flux density estimates, we derived a
spectral index $\alpha_{\rm PWN}=-0.18\pm 0.04$ and $\alpha_{\rm
shell}=-0.68\pm 0.07$. From the combination of the
radio data with X-ray data, a spectral break is found near $\nu\sim 2.4 \times 10^{12}$ Hz.
The total radio PWN luminosity is
L$_{\rm radio}=1.2\times10^{35}$ erg s$^{-1}$ when 
a distance of 8.5 kpc is adopted. By
assuming equipartition between particle and magnetic energies, we
estimate a nebular magnetic field B$ = 56~ \mu$G. The associated 
particle energy turns out to be U$_{\rm part}=5 \times 10^{47}$ erg and the
magnetic energy U$_{\rm mag}=2 \times 10^{47}$ erg.  The high
ratio between magnetic and particles flux energy density suggests that
the pulsar wind just started to become particle dominated.
 Based on an 
empirical relation between X-ray luminosity and pulsar energy loss rate, 
and the comparison with the calculated total energy, a lower limit of 1100 yr 
is derived for the age of this PWN.

}

\maketitle

\section{Introduction}
Radio composite supernova remnants
(SNRs)  consist of a shell and a spectrally distinct inner nebula,
presumably a pulsar wind nebula (PWN), powered by the wind of
relativistic electron/positron pairs from a central pulsar. Only in a
few cases, however, has the central pulsar been detected \citep[see][~for a review]{kaspihelfand02}.

Recent observations of several composite SNRs carried out with {\it
Chandra} X-ray Observatory have resolved out on arcsec scales complex 
structures in the
interior of several PWNe. These structures
include toroids, axial bipolar jets, wisps, etc.
\citep[][~etc.]{weiss00,helfand01,roberts03}. Images in the
different spectral domains are essential to understanding the physics
of PWNe.
Particularly, the radio emission depends on the history of the nebula
and represents the combination of the efficiency of the pulsar in
providing accelerated particles and magnetic fields, and the expansion
history. The expansion history, in turn, depends on the density and
geometry of
the medium that confines the relativistic particles and fields (i.e.
the interior of the SNR, that includes stellar ejecta and the presence
of forward and reverse shocks).  The detailed analysis of the geometry
and structure of the PWN and the
parent SNR, can shed light on the coupling mechanisms between the
neutron
star, the relativistic wind nebula and the surrounding SNR plasma.

G0.9+0.1 (RA= 17\hh 47\mm 21\ss, dec= -28\d 09\m,  J2000) is a composite 
SNR located in the direction of the Galactic
center and at about the same distance (assumed through this paper to
be   8.5 kpc).  It is characterized by a bright,
centrally condensed synchrotron nebula, approximately 2$^\prime$ 
in size, and a weak surrounding radio shell, about 8$^\prime$ in size, 
for which radio spectral indices $\alpha_{\rm core}\sim-0.12$ and 
$\alpha_{\rm shell}\sim-0.6$ (where S $\propto \nu^\alpha$), have 
been proposed for the core and shell respectively \citep{helfbec87,larosa00}.

In the X-rays domain, the first detection was reported by  
\citet{helfbec87} based on IPC-{\it Einstein} observations, who concluded that
the observed flux could come either from the compact core or from a
combination of core and part of the bright limb of the shell of G0.9+0.1. 
The  core X-ray emission  was detected  by \citet{mere98}
 using {\it BeppoSAX} satellite.
\citet{sidoli00} later confirmed these results 
on the basis of better quality data.
These early detections are indicative of the presence of a young
neutron star powering the nebula, although no 
coherent pulsations are found.
 
\citet{gaensler01} presented the results of 35 ksec ACIS
 {\it Chandra} observations of the PWN, between 0.5 and 8.0 keV. 
From these images, the authors identify 
a faint semicircular arc and a jet-like feature that define a 
symmetry axis, which they interpret  as evidence of a torus of emission in the
pulsar's
equatorial plane and a jet directed along the pulsar spin axis. No
X-ray emission is detected in correspondence with the radio-shell nor
its interior. 
Based on these observations the authors propose that the  
hard point-like X-ray source CXOU J174722.8-280915 detected
at energies above 3 keV, is the best candidate for a central pulsar
that would be powering the inner nebula. This point source has a
rather low ratio of magnetospheric pulsar emission
to surrounding nebular emission, with a  luminosity that amounts 
only 0.5 \% of the total PWN luminosity in the energy 
range 2-10 keV.

\citet{porquet03} carried out an X-ray study
of the PWN within G0.9+0.1 using the {\it XMM-Newton}
EPIC-MOS and EPIC-PN cameras. The images obtained in the energy band
1.5-12.0 keV show an amorphous nebula with a bright maximum towards
the east surrounded by extended diffuse emission. At the spatial 
resolution of {\it XMM-Newton} (8\arcsec),
the arc and jet-like features noticed by \citet{gaensler01} are not
obvious.  The X-ray spectrum within the PWN softens 
from the core to the
outskirts, consistent with synchrotron radiation losses of high energy
electrons as they diffuse through the nebula. The {\it XMM-Newton}  
study also reveals spectral variations across 
the ``arc-like feature''
identified by \citet{gaensler01}, with the eastern part of the arc 
having clear indications of a very hard photon index ($\Gamma \sim 1.0$),
opposite to the western part with  a very soft spectrum
($\Gamma \sim 3.2$). 

\citet{aharonian05} reported the detection, for the
first time, of gamma-ray emission in the direction of G0.9+0.1 
at energies greater than 100 GeV at
a level of significance of 13 $\sigma$. 
The very high energy gamma-rays, discovered using the H.E.S.S instrument, 
appear to
originate in the pulsar wind nebula. The photon spectrum is compatible with a
power law with photon index $\Gamma$ = 2.4.

\begin{figure}
   \centering
  \includegraphics[width=8 cm]{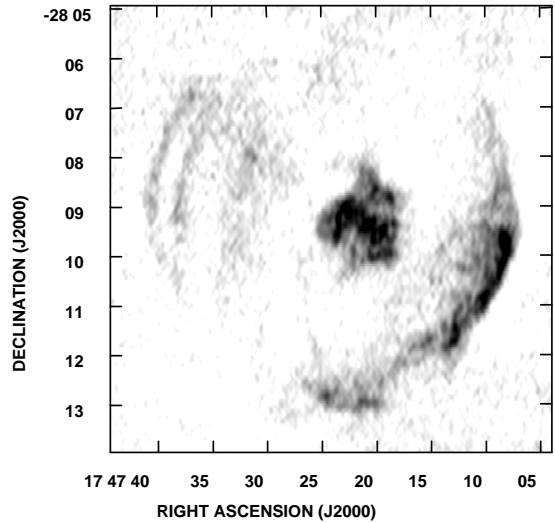}
      \caption{The  SNR G0.9+0.1 at 330 MHz as taken from \citet{nord04}.}
\label{fig:radio90cm}
   \end{figure}

In radio wavelengths, G0.9+0.1 is prominent at 57.5 MHz and 80 MHz
\citep{larosa85}. It has also been observed at 843 MHz \citep{gray94},
at 330 MHz as a part of the high-resolution imaging of the 
Galactic Center region  \citep{nord04} (Fig.~\ref{fig:radio90cm}), and at 1.5 GHz and
4.8 GHz  \citep{helfbec87}.

This paper attempts to analyze the morphology and spectral properties of 
the PWN in G0.9+0.1 in the radio range, based on new high-resolution
radio images obtained from observations at 4.8 GHz ($\lambda$ 6 cm) 
and at 8.4 GHz ($\lambda$ 3.6 cm) carried out
with the Australia Telescope Compact
Array\footnote{The Australia Telescope Compact Array (/ Parkes
telescope / Mopra telescope / Long Baseline Array) is part of the
Australia Telescope which is funded by the Commonwealth of Australia
for operation as a National Facility managed by CSIRO.} and from reprocessed archival VLA\footnote{The VLA of the
National Radio Astronomy Observatory is a facility of the NSF operated
under cooperative
agreement by the Associated Universities Inc.}  data at 1.4 GHz
($\lambda$  20 cm). Also the
X-ray emission associated with this nebula has been re-analyzed
including new {\it XMM-Newton} data and reprocessed {\it Chandra}
observations.

\section{Observations}

\subsection{Radio data at 3.6 cm, 6 cm and 20 cm}

The radio continuum emission of G0.9+0.1 was simultaneously 
observed at $\lambda$ 6 cm and  3.6 cm using the 
Australia Telescope 
Compact Array (ATCA) during 12 hours on 15/16 January 2004. The array was
used in the 6B configuration, which records visibilities from baselines 
214 m to 6 km. A total bandwidth of 128 MHz split in 32 channels was used
for each frequency. 
The absolute flux density scale was
determined using  PKS B1934-638 as the primary amplitude and bandpass 
calibrator (assuming 
S$_{6 \mathrm {cm}} =$ 5.83 Jy and S$_{3.6 \mathrm {cm}} =$ 2.84 Jy).
Periodic observations of PKS B1729-37 
were used to correct for changes in gain and phase caused by receiver, 
local oscillator and atmospheric instabilities.  Image
processing at all frequencies was carried out  under {\sc miriad}
software package \citep{sault99}.
  
To improve the uv
coverage, the $\lambda$ 6 cm ATCA 
observations were combined in the uv plane with VLA archive
data acquired in the same radio band  with the interferometer operating in the
hybrid  DnC configuration 
(program AB254 observed July 19 and 22 1984, PI R. Becker).
Finally, single dish data at $\lambda$ 6 cm acquired with the Parkes 
64 m telescope (extracted from Parkes-MIT-NRAO Southern Survey, Condon
et al. 1991) were combined  with the 
interferometric data  using the {\sc miriad} task {\sc immerge}, that
linearly merges together two images with different resolutions after 
appropriately weighting
both data sets according to their respective primary beam shapes. 
This addition allowed us to recover all missing flux
density due to the lack of short spatial frequencies. 
The faint structures associated with the  outer SNR shell are however 
barely detectable because in addition to
their intrinsic faintness they  suffer from attenuation produced near
the primary beam edge. 
The astrometry in the final
image was checked with the 6 cm Catalog of Compact Radio Sources in
the Galactic Plane \citep{white05} which has rms positional errors
lower than 0\arcsec.67 and 0\arcsec.84 in RA and dec, respectively. 
The resulting synthesized beam and rms noise in the $\lambda$ 6 cm
image are listed in
Table~\ref{table:obs_radio} 
together with the observational parameters for the other analyzed 
wavelengths. It is worthwhile to note
that the new $\lambda$ 6 cm image of G0.9+0.1 improves in over 50 times the
noise level with respect to the previously published image at the same
frequency.

 To produce a high-fidelity image at $\lambda$ 3.6 cm, we searched for
more databases at this frequency  to improve the
uv coverage. The ATCA data were combined with VLA D-array archive data 
obtained at the same frequency (observed on March 25 2003,
program  AJ302, PI M. Rupen), 
after applying an appropriate calibration
factor. This image is only useful to investigate the central PWN since
the surrounding radio shell (with a size of $\sim$ 8\m) exceeds the
primary beam size of the used telescopes at this wavelength
($\sim$ 5\m).

A new $\lambda$ 20 cm image was produced from the combination of archival VLA 
A-configuration data (observed in July 3 1991, program AF209, PI
D. Frail) with data obtained in the 
VLA hybrid BnC  configuration ( observed in February 6 and 7 2004,
program AY147, PI F. Yusef-Zadeh). 
The A-array observations were carried out at 1465 and 1515 MHz, using 
1331+305  and 1751-253 as primary and secondary calibrators, 
respectively. The CnB observations were done at  1385 and 1465 MHz. In
this case  1328+307 and 1748-253 were used as 
primary flux density and secondary phase calibrator, respectively. 
Because of the different observing conditions, both databases were separately 
calibrated and cleaned  and later combined  using 
the task {\sc immerge} within {\sc miriad} software, after deciding 
appropriate overlapping ring in the uv plane. In spite of the 
 special care put in the cleaning process to mitigate the effects of the strong 
neighbour source Sgr B2 within the observed field, 
some residual striation could not be completely removed. 

\begin{table}[htdp]
\caption{Observational parameters of the radio data}

\begin{center}
\begin{tabular}{|c|l|l|l|}
\hline
$\lambda$  & Instrument  & Beam & noise\\
(cm) & and configuration & (~\s$ \times $\s~)  & (mJy/b)\\
\hline
3.6 & ATCA (6B),VLA (D) & 1.5$\times$0.8  & 0.05  \\
6  & ATCA (6B),VLA (DnC),SD &2.9$\times$1.6& 0.07  \\
20  &  VLA (A,CnB) & 2.5$\times$1.2  & 0.12 \\
\hline
\end{tabular}
\end{center}
\label{table:obs_radio}
\end{table}%

\subsection{X-ray band}

We have analyzed the two existing sets of  {\it XMM-Newton} data on the 
PWN G0.9+0.1. The first observation, performed in September 2000, was 
published in \citet{porquet03}. A second longer observation was carried 
out in March 2003. The medium filter was set for all cameras (MOS and PN) 
in the two observations. The data were processed using the Science 
Analysis System (SAS version  7.1). The periods of high particle 
background (associated with flares) were rejected. Table~\ref{table:obs_xmm} 
summarizes the available observations and exposure time for each of the 
EPIC cameras, before and after flare screening. Despite the loss of a 
significant fraction of the observing time in the second observation, 
the statistics is improved compared to our previous analysis 
\citep{porquet03}, allowing a better comparison with the new high spatial 
resolution radio data. 

\begin{table}[htdp]
\caption{Summary of the {\it XMM-Newton} observations used in this paper.}
\begin{center}
\begin{tabular}{|c|c|c|c|}
\hline
Obs-Id & Instrument & Time & Time after \\
 &  &  &  flare screening\\
\hline 
0112970201 & M1 & 17.2 ks & 16.5 ks \\
0112970201 & M2 & 17.2 ks & 16.2 ks \\
0112970201 & PN & 11.8 ks &   9.3 ks \\
0144220101 & M1 & 49.4 ks & 25.3 ks \\
0144220101 & M2 & 49.4 ks & 29.0 ks  \\
0144220101 & PN & 43.7 ks & 15.7 ks  \\
\hline 
\end{tabular}
\end{center}
\label{table:obs_xmm}
\end{table}%

We also reprocessed archival {\it Chandra} data, originally published 
by \citet{gaensler01}, to produce an X-ray image of the PWN at higher 
spatial resolution. The observation 
(Obs-id=1036, Seq-num=500102) was performed in October 2000 for an exposure 
time of 35 ks. The image of the PWN was produced in the 3-8 keV energy band. 
The tool {\sc smooth} from the {\sc SAS} was used to provide an adaptively smoothed image at a signal to noise of 10.

\section {Radio results}


\begin{figure}
\hbox{

   \psfig{figure=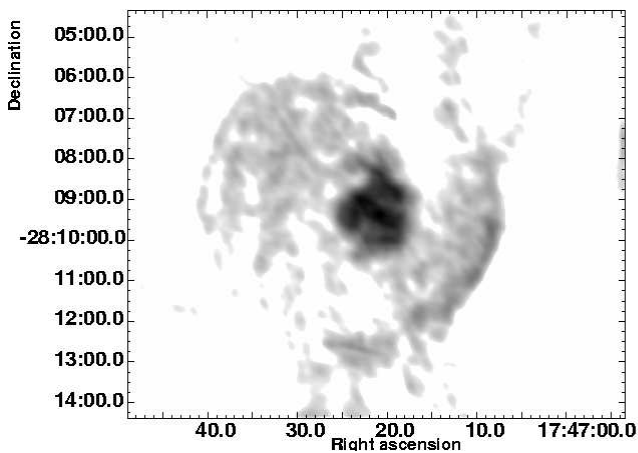,width= 10 cm}
        }
   \caption{ The SNR G0.9+0.1 at $\lambda$ 20 cm. The
angular resolution is 13\s.1 $\times$ 8\s.6 and the rms noise of 0.25
mJy beam$^{-1}$.
}
\label{fig:radio20cm}

    \end{figure}

Figure~\ref{fig:radio20cm} shows the new radio image of the SNR G0.9+0.1
obtained at $\lambda$ 20 cm based only on  the VLA CnB observations. The 
features are similar to those observed at $\lambda$ 90 cm  
(Fig.~\ref{fig:radio90cm}), with the brightest side of the surrounding
shell to the west and indication of  multiple faint arcs in the eastern half. 
This new sensitive image of G0.9+0.1 shows considerable diffuse
emission in the interior of the SNR. 

\subsection {Morphology of the PWN}

\begin{figure*}
\centering
   \psfig{figure=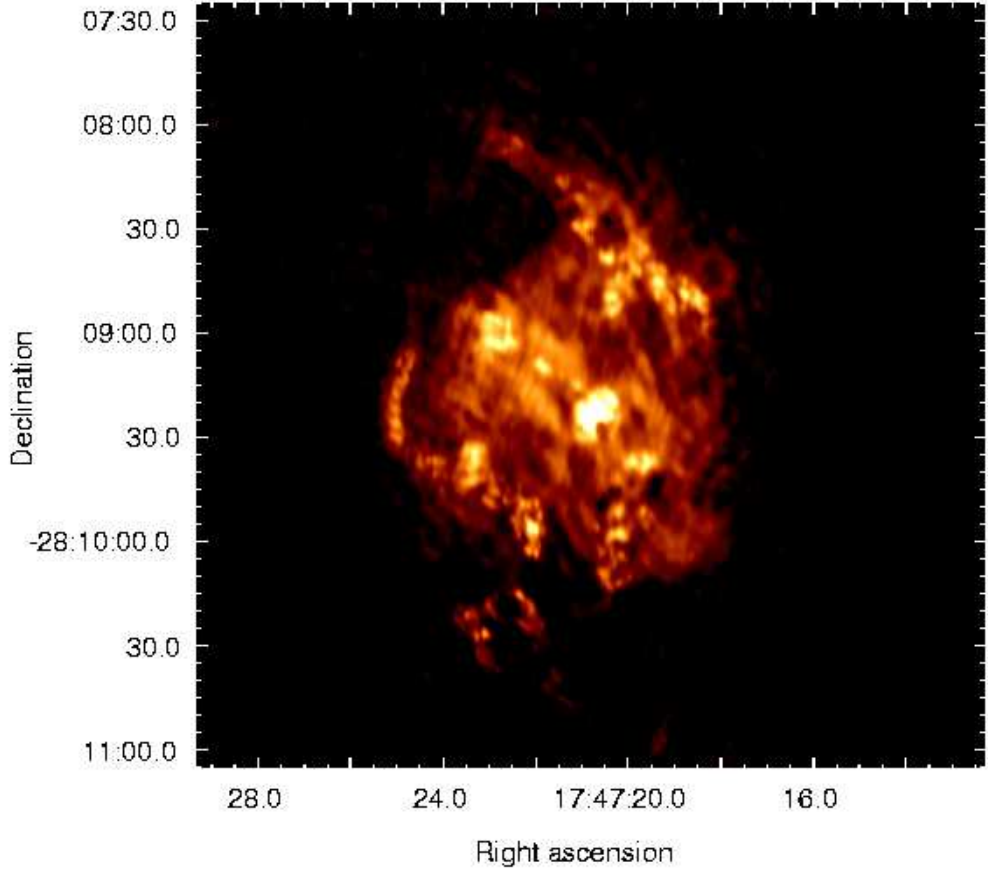,width=17 cm}
   \caption{ Color image of the central
PWN in the SNR G0.9+0.1 at $\lambda$ 6 cm. The angular resolution is 2\s.9
$\times$ 1\s.6 and the rms noise 0.07 mJy/b.}
\label{fig:6cm}
                                                                                
    \end{figure*}

\begin{figure*}
\centering
   \psfig{figure=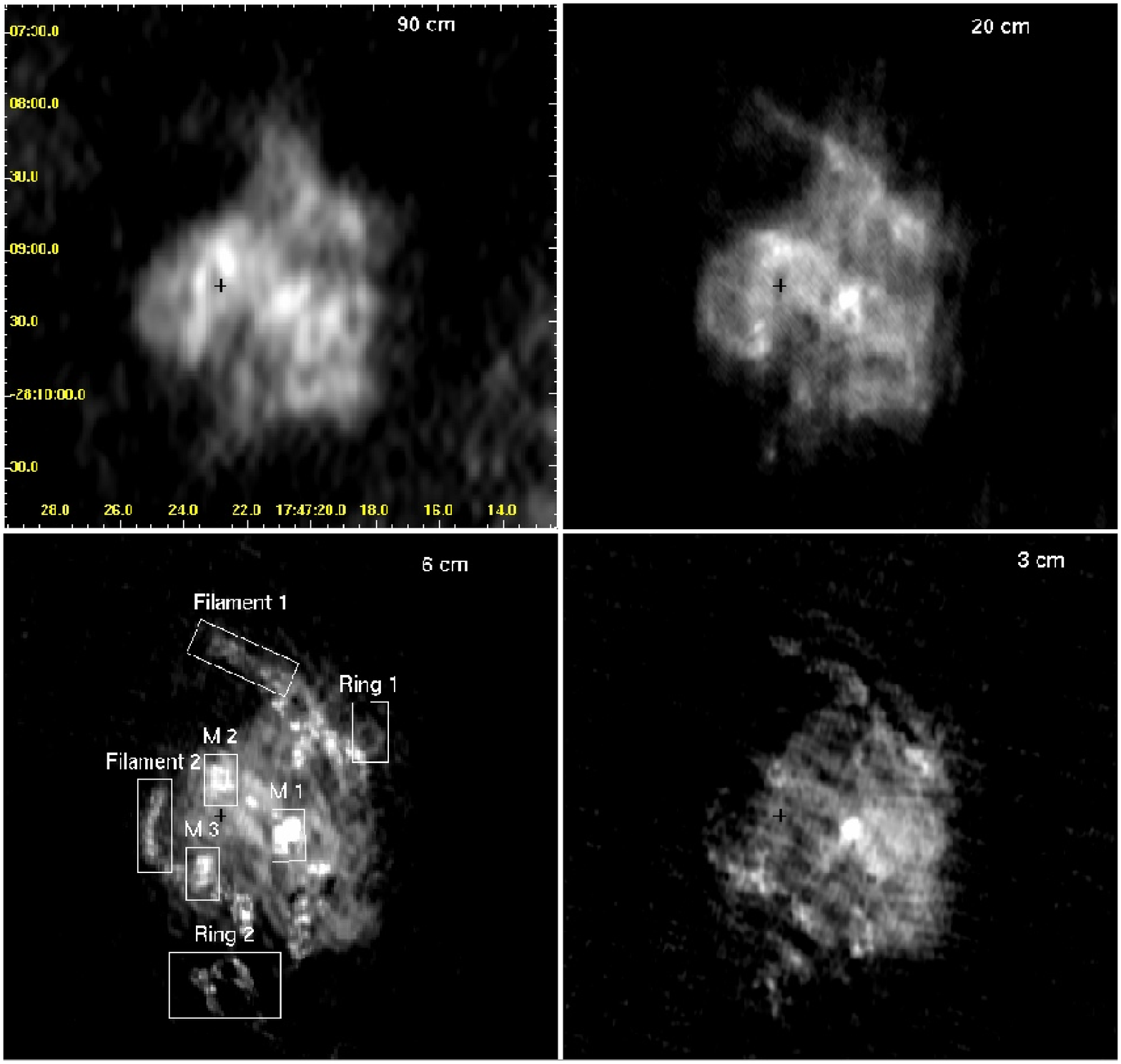,width=17 cm}
   \caption{ Grey-scale images of the central
PWN in the SNR G0.9+0.1 at 90, 20, 6 and 3.6 cm wavelengths. 
The black cross marks the position of the X-ray point
source \cxou. The different features individualized in the panel
corresponding to the 6 cm image are discussed in
the text.}
\label{fig:radiopwn}

    \end{figure*}

 In Fig.~\ref{fig:6cm} we display the new $\lambda$ 6 cm image of the PWN in
G0.9+0.1. Fig.~\ref{fig:radiopwn} includes  the  3.6,  6 and  
20 cm images of the PWN together with the 90 cm image taken from
\citet{nord04} for comparison. 

One of the major contributions of this work is the high-fidelity
representation attained in the 3.6 and 6 cm images, which have revealed
 that the radio synchrotron emission from
the confined wind  in G0.9+0.1 has a complex morphology 
with multiple small and large scale features, including enhancements
along filaments, bright knots and holes. The various conspicuous
features (individualized in the image at
$\lambda$ 6 cm in Fig.~\ref{fig:radiopwn}) are
present at all frequencies over the wide spectral range between 330 MHz (90 cm) and 8.6 GHz (3.6 cm),
although the respective brightness vary across the spectrum. 

The morphology is dominated by a  central band that runs
approximately from NE to SW mimicking at a larger scale the ``torus''
feature noticed in the {\it Chandra} image by \citet{gaensler01}. 
This band, that appears defining a
symmetry axis, has at least two main maxima (M1 and M2 in Fig.~\ref
{fig:radiopwn}), the brightest of which is the westernmost peak, M1.
 From the image at
$\lambda$ 6 cm, it looks as if this maximum is the result of the
overlap of two unresolved concentrations, of which only one (centered
near
17\hh47\mm22\ss.8,-28\d09\m00\s) remains visible at $\lambda$ 3 cm. 
The maximum M2, centered at 17\hh47\mm22\ss.8,-28\d09\m00\s, about
15\s~ north of the location of \cxou~ is resolved at $\lambda$ 6 and 3.6
cm, revealing that it is not compact, but hollow in the center.

Another interesting feature is the clumpy maximum M3, to the south of
the nebula. This maximum is apparently the termination of an 
 almost vertical filament that links
M2 with M3 (more clearly seen in the images at $\lambda$ 90 and 20
cm) and will be discussed below in connection with the X-ray emission. 
At the periphery of the wind nebula, four striking
features can be recognized. The northern
border  terminates  in a 
curious  filament pointing to the east 
(named Filament 1 in Fig.~\ref{fig:radiopwn}), 
while the southern limb also has a similar 
narrow filament on the eastern corner, but in this case the extension
points to the north along constant RA $\sim$ 17\hh47\mm25\ss (Filament
2). In addition, near the north-western corner a curious small synchrotron 
circular ring is present (Ring 1 in the $\lambda$ 6 cm image,
centered  near 17\hh47\mm18\ss, -28\d08\m40\s) and close to the
southern border of the nebula but detached from it, a set of 
short filaments can be observed  around $\sim $ 17\hh47\mm23\ss~,
-28\d10\m30\s~ forming another incomplete ring-like feature (Ring 2). 
The appearance of Filaments 1 and 2, protruding from 
the PWN, resemble the filaments observed in the Crab Nebula \citep{hester96}, 
and are likely to be originated 
in magnetic Rayleigh-Taylor instabilities at the interface between the 
expanding PWN and its surrounding SNR.
 
 Our sensitive new images do not reveal any point source
that could be interpreted as the radio  counterpart of \cxou, the faint
hard X-ray source proposed as the best candidate for a central pulsar
(indicated by a  black cross in Fig.~\ref{fig:radiopwn}). This is not
unexpected however, since a typical young radio pulsar distant
$\sim$ 8.5 kpc with an assumed  luminosity L$_{\rm 1.4 GHz} \sim$ 56 mJy
kpc$^2$ (like the median for
``high-luminosity'' young rotation-powered pulsars as estimated by
\citealp{camilo06}),  would have a flux density
S$_{\rm 1.4 GHz} \sim 0.08$ mJy, well below the sensitivity of the 1.4
GHz image. Moreover, to be detectable such a point-like source should
be brighter than the $\sim$ 70 mJ/beam nebular  emission in its vicinity.

\subsection{Flux density measurements and spectral study}

In the early paper  reporting the discovery
of G0.9+0.1, \citet{helfbec87} estimated the flux density of the 
PWN (called ``the core'' in
their paper) at $\lambda$ 6 and 20
cm on the basis of VLA observations. The calculations were carried out by
integrating the detected emission at both wavelengths 
within a square box 2\m~ on a side and, under some assumptions,
the authors derive S$_{\rm core-6 cm} =$ 4.16 Jy and 
S$_{\rm core-20 cm} =$ 4.15 Jy. For the
shell emission, at 6 cm the observational limitations 
(antennas shadowing limiting the VLA sensitivity for features on the
largest angular scales) are compensated by crudely adding a zero-spacing flux density 
to the uv data
before mapping. In this way the authors derive S$_{\rm shell-6 cm} \sim$ 7.9 Jy. For the estimates at
$\lambda$ 20 cm, the authors note that it is not possible to apply 
the same method than before
because of the presence of the bright source Sgr B2 at the edge of the
observed field, deriving under some considerations  
S$_{\rm shell-20 cm} \sim$ 14.4 Jy. Finally, these results were
combined with data collected from the literature between 408 MHz and
22 GHz to carry out a global spectral study. Here the authors reasonably 
note that 
because of the different beams used in the various multifrequency observations, 
it is apparent that some observers measured only the
core, i.e. the PWN alone, while others included some or all of the shell emission.
From this combination, \citet{helfbec87} obtain 
$\alpha_{\rm core}\sim -0.12$ and $\alpha_{\rm shell}\sim -0.6$, 
concluding that G0.9+0.1 is a composite SNR.  

Later \citet{larosa00} observed G0.9+0.1 at $\lambda$ 90 cm 
as a part of the Galactic Center study using the VLA (HPBW $\sim$ 43\s~ 
and  rms sensitivity 0.5 mJy beam$^{-1}$). In this work the authors 
report S$_{\rm shell-90
cm} \sim$ 16 Jy and S$_{\rm core-90 cm} \sim$ 3.5  Jy\footnote{ Note
that in Table 1 the authors list a flux density of 4.8 Jy for the core, 
while in the text the value mentioned for the same
component is 3.5 Jy, after correcting for diffuse background.} concluding, after
comparing with 20 cm VLA data, that $\alpha_{\rm core} \sim
+ 0.12$ and $\alpha_{\rm shell} \sim -0.77$, which confirms the
classification as a composite remnant suggested by Helfand \& Becker (1987).

Although it is beyond any doubt that the SNR G0.9+0.1 has two 
different spectral components, the PWN with a flat spectrum and the
shell with steeper spectrum, it is useful to revise the flux density
estimates based on  the new observations where the
contributions at all
spatial scales have been adequately considered and processed using
modern image reconstruction algorithms. Besides, the new images with
very good spatial resolution allow 
us to accurately determine  the areas considered to spatially integrate the 
flux density; also source brightness and background
contributions are  now estimated on the basis of very sensitive data. 

In Table~\ref{table:globalspix} we list the flux densities estimated from 
the new 3.6, 6 and 20 cm images.
The first column  lists the total SNR emission 
(that is, including contributions from the  external arcs, 
diffuse interior and core), the second  column corresponds to the 
core component alone, the third one has the same but with the 
underlying shell emission subtracted, and the fourth one lists the 
shell component alone.
The errors quoted in Table~\ref{table:globalspix} include the intrinsic  
noise as well as possible 
uncertainties introduced in 
the selection of the level of background contribution. 

At  $\lambda$ 3.6 cm only the flux density of the core component is
listed  because, as mentioned in $\S$2.1,  at this wavelength the shell can not be appropriately
mapped with a single pointing.
At $\lambda$ 6 cm,  the presence of diffuse  emission
associated
with the Galactic plane contaminates the SNR flux density estimates.
This effect was considered by subtracting a
background at a level that leaves the SNR G0.9+0.1
emission clearly detached from surrounding emission. Also, only lower
limit is listed for the flux density at $\lambda$ 6 cm of the whole SNR 
since, as mentioned in Section 2.1, at this
wavelength the outer shell contribution  is probably not completely
recovered. Though Table~\ref{table:globalspix} lists for completeness
the core flux
density before correcting for diffuse shell emission contribution, all
subsequent calculations are carried out based on the corrected flux
density.

\begin{table}[htdp]
\caption{Integrated radio flux densities}
                                                                                
\begin{center}
\begin{tabular}{|c|c|c|c|c|}
\hline
$\lambda$  &S$_{\rm SNR}$  & S$_{\rm PWN}$ & S$_{\rm
PWN}^{\rm corr}$&S$_{\rm shell}$ \\
(cm) &(Jy)  & (Jy)&(Jy)&(Jy)  \\
\hline
3.6  & &1.57$\pm$0.40& 1.35$\pm$0.50&  \\
6  & $\geq$6.90$\pm$0.20& 3.20$\pm$0.14& 1.45$\pm$0.23& $\geq$3.70$\pm$0.25 \\
20  & 11.50$\pm$0.70& 3.23$\pm$0.20  &1.72$\pm$0.30&  8.27$\pm$0.70 \\
\hline
\end{tabular}
\end{center}
\label{table:globalspix}

\end{table}

\begin{figure}
   \centering
  \includegraphics[width=8 cm]{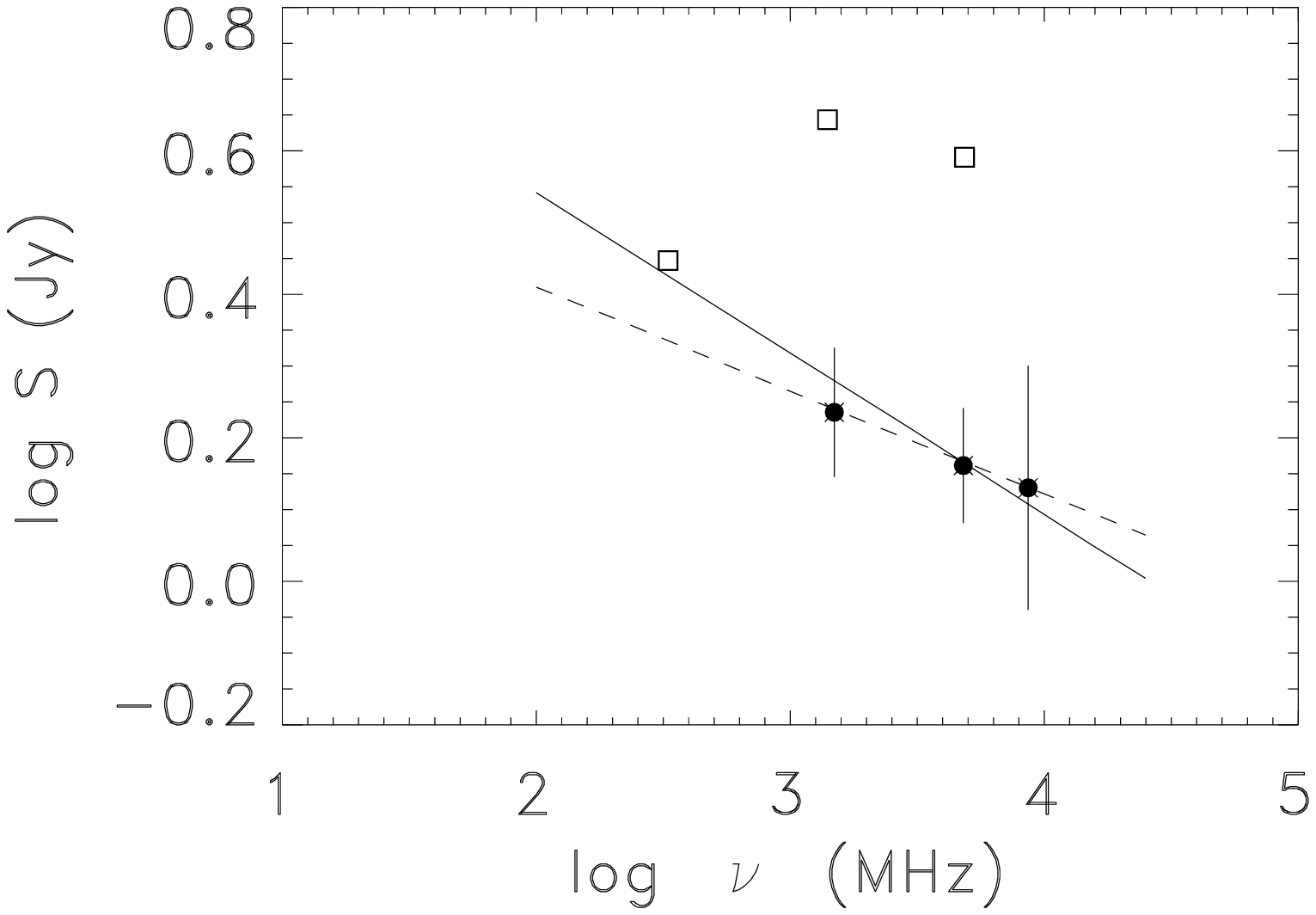}
      \caption{Radio spectrum of the PWN in G0.9+0.1. The data from
this work are shown by filled circles; data at 1400 and 4850 MHz from
\citet{helfbec87} and data at 330 MHz from \citet{larosa00} are
displayed as open squares. The best linear fit based on the new data plus
the 330 MHz (solid line) produces $\alpha=-0.22$. 
A linear fit to the new data alone (dashed line) produces
$\alpha=-0.14$. Data from
\citet{helfbec87} were not used in the fitting. An average
$\alpha=-0.18\pm0.04$ is adopted. } 
\label{fig:spectrum}
   \end{figure}

Figure~\ref{fig:spectrum} shows the global spectrum of the 
central PWN in G0.9+0.1 between 330 MHz and 8400 MHz together with the 
least square fits to our data alone (dashed line) and to our data 
plus \citealt{larosa00}'s 330 MHz
 data  (solid line).  The flux densities published by \citet{helfbec87} were not taken 
into account in this new fitting since, for the reasons mentioned above, 
they can be overestimated. In the case of including the 90 cm data, 
the PWN spectral index is $\alpha_{\rm PWN} =-0.22$, while based on 
20, 6 and 3.6 cm data alone, $\alpha_{\rm PWN} =-0.14$. In what follows 
we adopt for the PWN $\alpha_{\rm PWN} =-0.18 \pm 0.04$. 

For the shell component we have compared the 90 cm image with the new
20 cm image, obtaining $\alpha_{\rm shell} = -0.68\pm0.07$, in agreement 
with previous estimates.

The study of spatial variations of the spectrum across the PWN is a
sensitive tool for understanding the coupling between the fresh
relativistic electrons and magnetic fields constantly supplied by the
pulsar and the surrounding plasma. Therefore, based on the good
quality images obtained at three frequencies, we followed different
procedures to investigate possible 
spatial spectral variations across the PWN. 

 First the spectral study was
carried out by performing the direct comparison of the different images.
To assure that the range of spatial scales measured at
each frequency was perfectly matched, we applied an appropriate
uv tapering and reconstructed the interferometric images. In addition, 
to avoid positional
offsets, the images were aligned and interpolated to identical
projections (field center and pixel size). We repeated the ratio of the 
images, first at the angular 
resolution of the considered images and second, after
degrading the spatial resolution to 5\s $\times$ 5\s~ and 
to 10\s $\times$ 10\s~in order to minimize possible bias, 
like  small scale image artifacts, zero level 
differences,  background variations, etc., that could mask 
real variations. 
Contrary to what was shown in the X-rays domain by
\citet{porquet03} where the eastern side
of the PWN was found to have a clearly flatter (and harder) spectrum
than the southwestern half, we obtained that the radio spectral 
index distribution 
is practically featureless, with no particular 
morphology/spectrum correspondence or tendency. 

To investigate if the lack
of conspicuous spectral variations originated in the procedure, we repeated
the study using ``tomographic'' images \citep[see for example][]
{katzstone97}, a method where different spectral indices are tested and 
residuals with a spectrum flatter (or steeper) than the $\alpha_{\rm
test}$ 
are highlighted as darker (or lighter) features in an otherwise
uniform grey map. This method is very sensitive 
to fine-scale spectral index changes. Again the result
was that no clear departures from a mean spectral index of
$\alpha\sim -0.2$ are evident within the PWN. 
 To accurately trace small local variations in the spectral index, a set
of homogeneous observations acquired at different wavelengths using the same instrument and with analogous observing conditions, would be necessary.

\section{Radio/X-ray comparison}

The improved X-ray images of  G0.9+0.1 were used to compare with the 
radio brightness distribution. The full
comparison with X-ray images, however, lacks either better spatial resolution
(8\s~for XMM-Newton) or higher statistics (only 35 ks of Chandra
data). 

\begin{figure*}
  \centering
  \includegraphics[width=16cm]{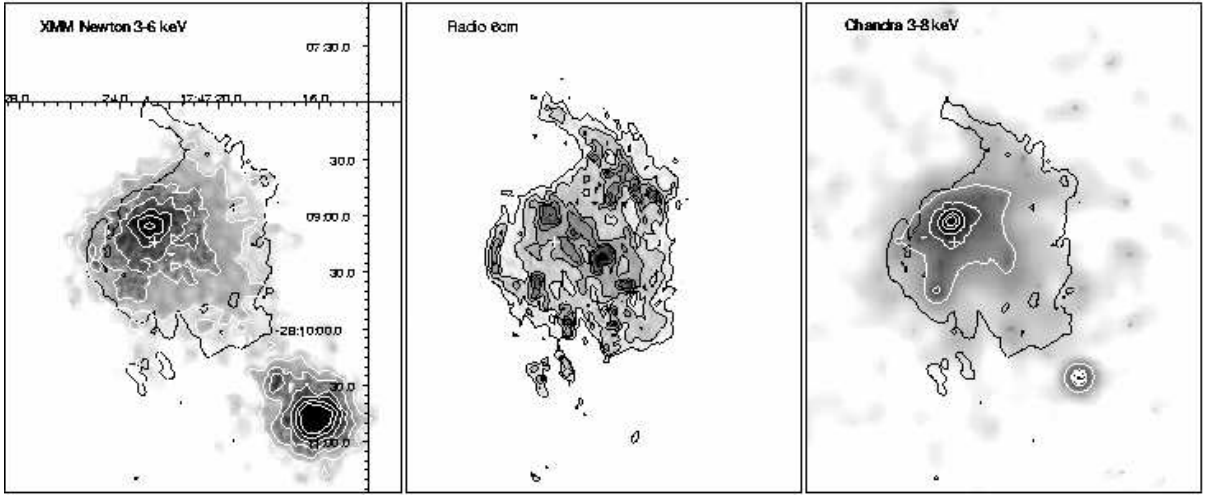}
\caption{Reprocessed {\it XMM-Newton} and {\it Chandra} X-ray images
displayed together with the 6 cm radio image. Both X-ray images have
X-ray contours (white)  and a radio contour (black) overlapped to
facilitate the identification of features and the multispectral
comparison. The location of the pulsar candidate \cxou~ is marked by 
a white plus sign.}
\label{fig:radiox}
 \end{figure*}

Figure~\ref{fig:radiox} shows the new  3-6 keV {\it XMM-Newton}
image  (left panel) and the 3-8 keV {\it Chandra} image 
 (right panel) compared with the radio image obtained at 6 cm (central panel to facilitate visual comparison with both X-ray datasets). 

It is significant the fact that the maxima in
the two spectral regimes do not coincide. The X-ray peak is
approximately 15\s~ southern of the radio maximum M2, while 
at the position of the brightest radio maximum, M1, only weak X-ray 
emission is detected.

The improved X-ray images confirm  that X-rays and radio emitting 
areas have a comparable extent. 
Good radio/X-ray correspondence can be
noticed between the termination of the ``jet-like''
feature noticed by \citet{gaensler01} in an approximate north-south
direction and interpreted as a jet directed along the pulsar spin axis, 
and  the radio maximum M3 
located exactly at the southern extreme of the X-ray feature.
Besides, as mentioned in $\S$3.1,  at $\lambda$ 90 cm and 
$\lambda$ 20 cm, the radio emission strikingly follows the shape and
extension of the entire X-ray jet.

\begin{figure}
   \centering
   \includegraphics[width=8 cm]{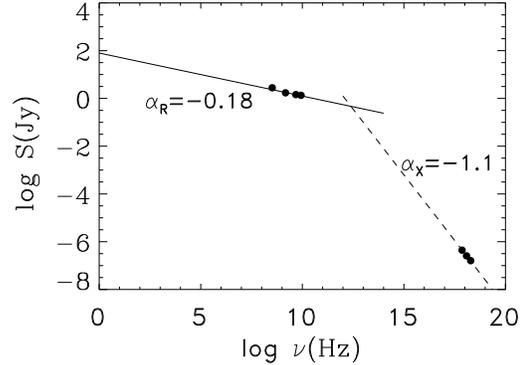}
   \caption{Spectrum of the PWN in G0.9+0.1 over the range 10$^8$
to 10$^{19}$ Hz, with a spectral index $\alpha_{\rm R}=-0.18$ in the
radio
regime and $\alpha_{\rm X}=-1.1$ as derived from least square fits to
the radio
and X-ray data, respectively.  The extrapolation of the X-ray
spectrum intersects that of the radio spectrum at the turnover
frequency
$\nu_{\rm b}=2.4\times10^{12}$ Hz.}

\label{fig:nubreak}
   \end{figure}

Based on the {\it XMM-Newton} data we have estimated the X-ray flux in 
: $2.14\times10^{12}$ erg cm$^{-2}$s$^{-1}$ between 2 and 4 keV,
$1.24\times10^{12}$ erg cm$^{-2}$s$^{-1}$ between 4 and 6 keV, and  
$1.56\times10^{12}$ erg cm$^{-2}$s$^{-1}$ in the 6 to 10 keV interval.
Following the procedure used for the radio data, we applied a least 
square fit to the X-ray data, obtaining a spectral index $\alpha_{\rm
X}=$ -1.1, which corresponds to a photon index $\Gamma= 1-\alpha =
2.1$, which equals the average value between the hardest spectrum derived by
\citet{porquet03} on the eastern side of the PWN ($\Gamma=1.0$) 
and the softest one ($\Gamma=3.2$) on the western part. In
Figure~\ref{fig:nubreak}  we display the radio synchrotron spectrum
shown in Fig.~\ref{fig:spectrum}  together with the X-ray data.
 From the intersection of the spectral fitting in
radio and in the X-ray domain, we can conclude that the spectrum
must break around $\nu_{\rm b}= 2.4\times10^{12}$ Hz.

\section{Energetics of the PWN}
                                     
Based on the estimated radio flux density,  break frequency $\nu_{\rm
b}$ and radio spectral index, we can calculate the radio luminosity 
associated with the PWN in G0.9+0.1 between
$\sim 10^7$ Hz and  $\nu_{\rm
b}$. A total L$_{\rm radio} \sim 1.2 \times 10^{35}$ erg s$^{-1}$, is
obtained. This radio luminosity can be compared to L$_{\rm X} \sim
4.7 \times 10^{34}$ erg s$^{-1}$ measured by \citet{porquet03} between
2 and 10 keV, after correcting for d=8.5 kpc. The luminosities ratio, 
a useful parameter because it is independent of the distance, turns out to be
L$_{\rm X}$/ L$_{\rm radio} \sim$ 0.4. 

It is of great interest to determine the energy requirements for the 
radio synchrotron emission associated with the PWN.
Energy is stored in the particles as well as in the magnetic field.
 Following  \citet{moffett75} we can express the total energy of the
source as
$$ {\rm U_T} = {\rm U_p} + {\rm U_m} = aALB^{-3/2} + VB^2/8\pi$$

where  L is the total radio luminosity
of the nebula, V its volume,
{\it a} represents the ratio between the energy in relativistic
electrons
and in energetic baryons, and is assumed to be 1,  while A depends on
the
lower an upper cutoff frequencies and can be estimated as
follows:
$$
A= {{{C_1}^{1/2}} \over {C_3}} {{(2 \alpha + 2)} \over {(2 \alpha +
1)}}
{{({\nu_1}^{\alpha+1/2} - {\nu_0}^{\alpha+1/2})} \over
{({\nu_1}^{\alpha+1} - {\nu_0}^{\alpha+1})}}
$$

The total energy as a function of the magnetic field $B$ attains its
minimum when the magnetic energy is approximately equal to the particle energy
(equipartition situation). It is usually assumed
that the magnetic field in the source is actually equal to the
value which gives minimum energy. The minimum total energy   
is  ${\rm U}_{\rm min} =0.50 (a A {\rm L})^{4/7} {\rm V}^{3/7} = 7
\times 10^{47} {\rm erg}$ and  the magnetic field for which equipartition holds can be calculated from: 
${\rm B (U}_{\rm min}) = 2.3 (a A {\rm L/V}) ^{2/7} = 56~ \mu{\rm G}$.
This value agrees with  the magnetic field strength 
early proposed by \citet{helfbec87} for
this PWN (though they assumed a higher $\nu_{\rm b}$), 
and is comparable to the value estimated for the 
PWN in the SNR W44 \citep{petre02}. \citet{aharonian05} estimated a  
magnetic field strength  of only 6~ $\mu$G for the PWN in G0.9+0.1, though 
this number is inferred from model fitting to the
broad-band spectral energy distribution, where the B field is a free 
parameter. 

Based on our calculated magnetic field, the  magnetic energy is 
${\rm U}_{\rm m} \sim 2 \times 10^{47} {\rm erg}$ and
the particle energy ${\rm U}_{\rm p} \sim 5 \times 10^{47} {\rm erg}$.   
The wind magnetization parameter (ratio 
between magnetic flux energy density to that in particles)
is then $\sigma \sim$ 0.4. Such a high value suggests that
 the wind just started to become particle dominated, though the
transition from $\sigma > 1$ to  $\sigma < 1$ is not yet clear
\citep[see][~for discussion]{arons98}. 

A rough estimate of the 
age of the PWN can be estimated from the comparison of  the  total 
energy with the rate of rotational energy loss of the pulsar. This
last parameter can be derived from the empirical relation between 
the X-ray luminosity
and $\dot {\rm E}$ proposed by
\citet{possenti02}:$$ {\rm log~ L}_{\rm X}=1.34~ {\rm log}~{\dot {\rm
E}}-15.3$$ 
From this relation, an energy loss rate $\dot {\rm E \sim} 2\times10^{37}$
erg s$^{-1}$ is derived, thus implying that the pulsar had to be
continuously injecting energy during, at least, $\sim$ 1100 yrs to create
the observed PWN.

\begin{figure}
   \centering
   \includegraphics[width=8 cm]{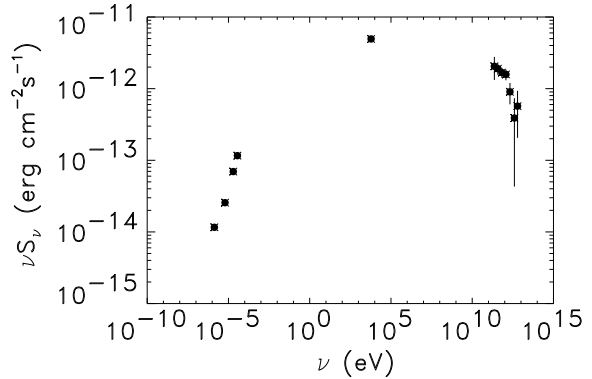}
      \caption{Spectral energy distribution of the PWN in G0.9+0.1
from radio to VHE gamma-ray, joining the new estimates in the radio
and X-ray bands with HESS measurements.} 
                                                                                
\label{fig:nusnu}
   \end{figure}

As a by-product of the new radio measurements, in Figure~\ref{fig:nusnu} 
we show the broadband spectral energy 
distribution (SED) of the PWN in G0.9+0.1 obtained after combining the  
radio data presented in this paper plus the 330 MHz data from
\citet{larosa00} together with the X-ray data and VHE HESS measurements
(from the public HESS database). 

\section{Conclusions}

This paper presents new high-resolution and high-sensitivity images
of the PWN in the SNR G0.9+0.1 obtained at different radio frequencies.
The study is complemented with reprocessed X-ray images based on {\it
XMM-Newton} and {\it Chandra} data. The new radio images have revealed
interesting structures in the nebula, like bright knots, rings
and elongated filaments which might be showing instability regions
at the sites where the expanding nebula interacts with the surrounding 
ejecta.  From the comparison of the radio images with the reprocessed 
X-ray images it is found that the X-ray emitting electrons largely fills the
volume delineated by the radio PWN. Also, the
new  detailed radio images have confirmed  the symmetry suggested by
the {\it Chandra} X-ray observations, with a bright central band
aligned with the X-ray ``toroidal'' feature and a narrow elongated
north-south structure that appear as the counterpart of the ``jet-like''
X-ray feature. These good radio/X-ray correspondences are, however,
accompanied by notable disagreements, the most important of which is
the separation observed between the radio and the X-ray maxima. 

Based on the new radio images, with contributions from all spatial
scales adequately recovered, we estimated the
multispectral flux densities and performed a spectral study. In 
Table~ \ref{table:summary} we summarize these results together with
other observed and derived
characteristic parameters of the PWN in G0.9+0.1. Our study 
revealed a quite uniform distribution of radio spectral index across
the nebula, with only small fluctuations around the mean value of
$\alpha_{\rm r}= -0.18$.

\begin{table}[htdp]
\caption{Characteristic parameters of the PWN in G0.9+0.1}
                                                                                
\begin{center}
\begin{tabular}{r l}
S$_{\rm 3.6 cm}=$ & 1.35$\pm$0.50 Jy\\
S$_{\rm 6 cm}=$  & 1.45$\pm$0.23 Jy\\
S$_{\rm 20 cm}=$  & 1.72$\pm$0.30 Jy\\
$\alpha_{\rm radio}=$& $-0.18\pm0.04$\\
L$_{\rm radio}\simeq$ & $ 1.2 \times 10^{35}$ erg s$^{-1}$\\
$\nu_{\rm b}\simeq$ & $2.4\times10^{12}$ Hz\\
B ~$\simeq$ & 56 $\mu$G\\
U$_{\rm particles}\simeq$ & 5 $\times 10^{47}$ erg\\
U$_{\rm magnetic}\simeq$ &  2 $\times 10^{47}$ erg\\
age $\geq$ &1100 yr\\
\end{tabular}
\end{center}
\label{table:summary}
\end{table}

From the combination of observations in the radio regime with X-ray
data we  traced a broadband spectrum which suggests a spectral break
at $\nu_{\rm b} \approx 2.4\times10^{12}$ Hz. On the basis of this information,
together with the observed luminosities
and the assumption of equipartition between particles energy and
Poynting vector energy, we investigated the energetics and the
magnetic field  in the nebula.

\begin{acknowledgements}
We are very grateful to E. Reynoso and to A. Green who participated in
the data acquisition and first stages of this work.  
We thank the referee, Professor David Helfand, for his useful comments 
that improved the manuscript. We acknowledge
M. Nord for providing us with the 330 MHz image used for this study.
This research was carried out within the framework of the ECOS-Sud
France-Argentina exchange program. The research has been partially
funded by grants CONICET PIP 6433, UBACYT A055/04
and ANPCYT-PICT 03-14018 of Argentina. This work is based on
observations done with {\it XMM-Newton}, an ESA science mission with
instruments and contributions directly funded by ESA Member States and
the US (NASA).

\end{acknowledgements}


\bibliographystyle{aa}  
\bibliography{g09-references.bib}

\begin{thebibliography}{23}
\expandafter\ifx\csname natexlab\endcsname\relax\def\natexlab#1{#1}\fi

\bibitem[{{Aharonian} {et~al.}(2005){Aharonian}, {Akhperjanian}, {Aye},
  {Bazer-Bachi}, {Beilicke}, {Benbow}, {Berge}, {Berghaus}, {Bernl{\"o}hr},
  {Boisson}, {Bolz}, {Borgmeier}, {Braun}, {Breitling}, {Brown}, {Bussons
  Gordo}, {Chadwick}, {Chounet}, {Cornils}, {Costamante}, {Degrange},
  {Djannati-Ata{\"i}}, {O'C.~Drury}, {Dubus}, {Ergin}, {Espigat}, {Feinstein},
  {Fleury}, {Fontaine}, {Funk}, {Gallant}, {Giebels}, {Gillessen}, {Goret},
  {Hadjichristidis}, {Hauser}, {Heinzelmann}, {Henri}, {Hermann}, {Hinton},
  {Hofmann}, {Holleran}, {Horns}, {de Jager}, {Jung}, {Kh{\'e}lifi}, {Komin},
  {Konopelko}, {Latham}, {Le Gallou}, {Lemi{\`e}re}, {Lemoine}, {Leroy},
  {Lohse}, {Marcowith}, {Masterson}, {McComb}, {de Naurois}, {Nolan},
  {Noutsos}, {Orford}, {Osborne}, {Ouchrif}, {Panter}, {Pelletier}, {Pita},
  {P{\"u}hlhofer}, {Punch}, {Raubenheimer}, {Raue}, {Raux}, {Rayner},
  {Redondo}, {Reimer}, {Reimer}, {Ripken}, {Rob}, {Rolland}, {Rowell},
  {Sahakian}, {Saug{\'e}}, {Schlenker}, {Schlickeiser}, {Schuster}, {Schwanke},
  {Siewert}, {Sol}, {Steenkamp}, {Stegmann}, {Tavernet}, {Terrier},
  {Th{\'e}oret}, {Tluczykont}, {Vasileiadis}, {Venter}, {Vincent}, {Visser},
  {V{\"o}lk}, \& {Wagner}}]{aharonian05}
{Aharonian}, F., {Akhperjanian}, A.~G., {Aye}, K.-M., {et~al.} 2005, \aap, 432,
  L25

\bibitem[{{Arons}(1998)}]{arons98}
{Arons}, J. 1998, Memorie della Societa Astronomica Italiana, 69, 989

\bibitem[{{Camilo} {et~al.}(2006){Camilo}, {Ransom}, {Gaensler}, {Slane},
  {Lorimer}, {Reynolds}, {Manchester}, \& {Murray}}]{camilo06}
{Camilo}, F., {Ransom}, S.~M., {Gaensler}, B.~M., {et~al.} 2006, \apj, 637, 456

\bibitem[{{Gaensler} {et~al.}(2001){Gaensler}, {Pivovaroff}, \&
  {Garmire}}]{gaensler01}
{Gaensler}, B.~M., {Pivovaroff}, M.~J., \& {Garmire}, G.~P. 2001, \apjl, 556,
  L107

\bibitem[{{Gray}(1994)}]{gray94}
{Gray}, A.~D. 1994, Proceedings of the Astronomical Society of Australia, 11,
  79

\bibitem[{{Helfand} \& {Becker}(1987)}]{helfbec87}
{Helfand}, D.~J. \& {Becker}, R.~H. 1987, \apj, 314, 203

\bibitem[{{Helfand} {et~al.}(2001){Helfand}, {Gotthelf}, \&
  {Halpern}}]{helfand01}
{Helfand}, D.~J., {Gotthelf}, E.~V., \& {Halpern}, J.~P. 2001, \apj, 556, 380

\bibitem[{{Hester} {et~al.}(1996){Hester}, {Stone}, {Scowen}, {Jun},
  {Gallagher}, {Norman}, {Ballester}, {Burrows}, {Casertano}, {Clarke},
  {Crisp}, {Griffiths}, {Hoessel}, {Holtzman}, {Krist}, {Mould}, {Sankrit},
  {Stapelfeldt}, {Trauger}, {Watson}, \& {Westphal}}]{hester96}
{Hester}, J.~J., {Stone}, J.~M., {Scowen}, P.~A., {et~al.} 1996, \apj, 456, 225

\bibitem[{{Kaspi} \& {Helfand}(2002)}]{kaspihelfand02}
{Kaspi}, V.~M. \& {Helfand}, D.~J. 2002, in Astronomical Society of the Pacific
  Conference Series, 3

\bibitem[{{Katz-Stone} \& {Rudnick}(1997)}]{katzstone97}
{Katz-Stone}, D.~M. \& {Rudnick}, L. 1997, \apj, 479, 258

\bibitem[{{La Rosa} \& {Kassim}(1985)}]{larosa85}
{La Rosa}, T.~N. \& {Kassim}, N.~E. 1985, \apjl, 299, L13

\bibitem[{{La Rosa} {et~al.}(2000){La Rosa}, {Kassim}, {Lazio}, \&
  {Hyman}}]{larosa00}
{La Rosa}, T.~N., {Kassim}, N.~E., {Lazio}, T.~J.~W., \& {Hyman}, S.~D. 2000,
  \aj, 119, 207

\bibitem[{{Mereghetti} {et~al.}(1998){Mereghetti}, {Sidoli}, \&
  {Israel}}]{mere98}
{Mereghetti}, S., {Sidoli}, L., \& {Israel}, G.~L. 1998, \aap, 331, L77

\bibitem[{{Moffet}(1975)}]{moffett75}
{Moffet}, A.~T. 1975, {Strong Nonthermal Radio Emission from Galaxies}
  (Galaxies and the Universe), 211--+

\bibitem[{{Nord} {et~al.}(2004){Nord}, {Lazio}, {Kassim}, {Hyman}, {LaRosa},
  {Brogan}, \& {Duric}}]{nord04}
{Nord}, M.~E., {Lazio}, T.~J.~W., {Kassim}, N.~E., {et~al.} 2004, \aj, 128,
  1646

\bibitem[{{Petre} {et~al.}(2002){Petre}, {Kuntz}, \& {Shelton}}]{petre02}
{Petre}, R., {Kuntz}, K.~D., \& {Shelton}, R.~L. 2002, \apj, 579, 404

\bibitem[{{Porquet} {et~al.}(2003){Porquet}, {Decourchelle}, \&
  {Warwick}}]{porquet03}
{Porquet}, D., {Decourchelle}, A., \& {Warwick}, R.~S. 2003, \aap, 401, 197

\bibitem[{{Possenti} {et~al.}(2002){Possenti}, {Cerutti}, {Colpi}, \&
  {Mereghetti}}]{possenti02}
{Possenti}, A., {Cerutti}, R., {Colpi}, M., \& {Mereghetti}, S. 2002, \aap,
  387, 993

\bibitem[{{Roberts} {et~al.}(2003){Roberts}, {Tam}, {Kaspi}, {Lyutikov},
  {Vasisht}, {Pivovaroff}, {Gotthelf}, \& {Kawai}}]{roberts03}
{Roberts}, M.~S.~E., {Tam}, C.~R., {Kaspi}, V.~M., {et~al.} 2003, \apj, 588,
  992

\bibitem[{{Sault} \& {Killeen}(1999)}]{sault99}
{Sault}, B. \& {Killeen}, N. 1999, {Miriad Users Guide,
  http://www.atnf.csiro.au/computing/software/miriad}

\bibitem[{{Sidoli} {et~al.}(2000){Sidoli}, {Mereghetti}, {Israel}, \&
  {Bocchino}}]{sidoli00}
{Sidoli}, L., {Mereghetti}, S., {Israel}, G.~L., \& {Bocchino}, F. 2000, \aap,
  361, 719

\bibitem[{{Weisskopf} {et~al.}(2000){Weisskopf}, {Hester}, {Tennant}, {Elsner},
  {Schulz}, {Marshall}, {Karovska}, {Nichols}, {Swartz}, {Kolodziejczak}, \&
  {O'Dell}}]{weiss00}
{Weisskopf}, M.~C., {Hester}, J.~J., {Tennant}, A.~F., {et~al.} 2000, \apjl,
  536, L81

\bibitem[{{White} {et~al.}(2005){White}, {Becker}, \& {Helfand}}]{white05}
{White}, R.~L., {Becker}, R.~H., \& {Helfand}, D.~J. 2005, \aj, 130, 586

\end{thebibliography}
\IfFileExists{\jobname.bbl}{}
{\typeout{}
\typeout{****************************************************}
\typeout{****************************************************}
\typeout{** Please run "bibtex \jobname" to optain}
\typeout{** the bibliography and then re-run LaTeX}
\typeout{** twice to fix the references!}
\typeout{****************************************************}
\typeout{****************************************************}
\typeout{}
}

\end{document}